# Physics behind the Debye temperature

József Garai [a)]

*Department of Mechanical and Materials Engineering, Florida International University, Miami, USA*

Textbooks introduce the Debye temperature to simplify the integration of the heat capacity. This approach gives the impression that the Debye temperature is a parameter which makes the integration more convenient. The Debye frequency cut occurs when the wavelength of the phonon frequency reaches the size of the smallest unit of the lattice which is the length of the unit cell. At frequencies higher than the cut off frequency the "lattice" unable to "see" the vibration because the wavelength of the vibration is smaller than the basic unit of the atomic arrangement; therefore, the vibration becomes independent from the lattice. The Debye cut off frequency or temperature separates the collective thermal lattice vibration from the independent thermal lattice vibration. The experimental data of highest packing monoatomic arrangements is used to calculate the wavelength of the Debye cut off frequency. The calculated values agree well with the unit cell parameters.

**I. INTRODUCTION**

In solid phase, by assuming small displacement from the equilibrium position, atoms execute simple vibrational motion in the x, y, and z direction which is called lattice vibration. Assuming linear elasticity the internal energy of a system [U] can be calculated as:

$$U = \sum_{i=1}^{N}\sum_{\alpha=1}^{3}\left(\frac{p_{i\alpha}^{2}}{2m_{i}} + \frac{1}{2}\kappa_{i}\xi_{i\alpha}^{2}\right) \qquad (1)$$

where $\alpha = 1, 2,$ or $3$, p is the momentum of the particle, m is the mass of the particle, $\kappa$ is the

force constant, and N is the number of the vibrating units. The variable $\xi$ is defined as:

$$\xi_{i\alpha} \equiv x_{i\alpha} - x_{i\alpha}^{(0)} \tag{2}$$

where $x^{(0)}$ is the equilibrium position of the particle[1]. If the temperature is high enough then classical physics is applicable and Eq.(1) can be used to determine the internal energy of the system. Applying the equipartition theory gives the internal energy

$$U = 3N\left[\left(\frac{1}{2}\right)k_B T \times 2\right] = 3nN_A k_B T = 3nRT, \tag{3}$$

where $k_B$ is the Boltzmann constant, T is the absolute temperature, R is the universal gas constant, n is the number of mols and $N_A$ is the Avogadro number. The molar heat capacity at constant volume $[c_V]$ is

$$c_V = \frac{1}{n}\frac{\partial U}{\partial T} = \frac{\partial}{\partial T}(3N_A k_B T) = 3R = 24.94 J/mol. \tag{4}$$

This theoretical value of the molar volume heat capacity is known as the DuLong-Petit law[2]. At room temperatures Eq. (4) is in excellent agreement with experiments; however, at low temperatures significant deviation is observed. Einstein proposed that the independent harmonic vibration of the atoms around their equilibrium position must be quantized[3] in accordance with Planck's suggestion[4,5].

Both the Dulong-Petit and Einstein models assume independent atomic vibration. Born and Von Karman[6] proposed that the bonding in solid phase prevents independent atomic vibration and that the vibration should be collective lattice oscillation. If the particles are interacting then the potential energy of the particle in the crystal depends on the distance from its neighbor and Eq.(1) must be modified as:

$$U = \sum_{i=1}^{N}\sum_{\alpha=1}^{3}\frac{p_{i\alpha}^2}{2m_i} + \frac{1}{2}\sum_{i=1,j=1}^{N}\sum_{\alpha=1}^{3}\kappa_{ij}(\xi_{i\alpha} - \xi_{j\alpha})^2 \tag{5}$$

The three dimensional bonding interactions are complex. In order to overcome on the complexities Debye simplified the problem[7] by assuming that the velocity of the sound in solid $[v_s]$ is constant for all the frequencies leading to $\sigma_D(\omega) \propto \omega^2$. The spectral distribution then can be defined as:

$$\sigma_D(\omega) \begin{cases} \sigma_C(\omega) & \text{for } \omega \leq \omega_D \\ 0 & \text{for } \omega > \omega_D \end{cases} \quad (6)$$

where $\omega_D$ is the Debye frequency. The distribution is terminated at the point when the number of vibration becomes equal with the number of degrees of freedom. Introducing the Debye function

$$f = 3\left(\frac{T}{\theta_D}\right)^3 \int_0^{x_D} \frac{x^4 e^x}{(e^x - 1)^2} dx \quad (7)$$

and

$$x = \frac{h\omega}{2\pi k_B T} \quad \text{and} \quad x_D = \frac{h\omega_D}{2\pi k_B T} = \frac{T}{T_D} \quad (8)$$

where $T_D$ the Debye temperature defined as:

$$T_D \equiv \frac{h\omega_D}{2\pi k_B}. \quad (9)$$

The molar heat capacity at constant volume can be calculated as:

$$c_V = 3Rf = 9N_A k_B \left(\frac{T}{\theta_D}\right)^3 \int_0^{\theta_D/T} \frac{x^4 e^x}{(e^x - 1)^2} dx. \quad (10)$$

Equation (10) has to be evaluated numerically[8].

The introduction of the Debye model is part of the standard curriculum in any solid state physics course[9, 10] and has been covered from different aspects in this journal[11-16]. Textbooks define the Debye temperature in the form of Eq. (9) as part of the integration of the heat capacity[1,9, 10, 17]. This approach gives the impression that the Debye temperature is a parameter which makes the integration easier. The Debye temperature has very important physical meaning which should be clearly explained to students. An example of such explanation is presented.

## II. THEORY

In solid phase the lattice vibrations are quantized and can be described by quasi-particles, phonons. It is assumed that the phonons represent a normal mode vibration thus all parts of the lattice vibrate with the same frequency. This quantum mechanical approach can be modeled as a particle in a box.

Assuming that the particle is in a cubic box and that the length of the box is L then the wavelength $[\lambda_n]$ of the phonon vibration is

$$\lambda_n = \frac{2L}{n} \tag{11}$$

where n is a positive integer. The frequency of the vibrating phonons $[\nu_n]$ can be determined from the speed of the sound inside the solid

$$\nu_n = \frac{\upsilon_m}{\lambda_n} = \frac{n\upsilon_m}{2L}. \tag{12}$$

where $[\upsilon_m]$ represents the mean acoustic velocity calculated from the longitudinal $[\upsilon_p]$, and transverse $[\upsilon_S]$ sound velocities as:

$$\upsilon_m \equiv 3^{\frac{1}{3}}\left(\frac{2}{\upsilon_S^3} + \frac{1}{\upsilon_p^3}\right)^{-\frac{1}{3}}. \tag{13}$$

The energy of the phonon is

$$E_n = h\nu_n \tag{14}$$

where h is a Planck's constant. Substituting Eq. (12) into Eq. (14) gives the energy of the phonon as:

$$E_n = h\nu_n = \frac{h\upsilon_m}{\lambda_n} = \frac{nh\upsilon_m}{2L} \tag{15}$$

In three dimensional spaces the phonon energy is the sum of its dimensional components

$$E_n = \sqrt{E_{nx}^2 + E_{ny}^2 + E_{nz}^2} = \sqrt{\left(\frac{h\upsilon_m}{2L}\right)^2 \left(n_x^2 + n_y^2 + n_z^2\right)}. \tag{16}$$

The total energy of the sonic vibrating system is then

$$U = \sum_{n_x}\sum_{n_y}\sum_{n_z} E_n \overline{N}(E_n). \tag{17}$$

where $\overline{N}(E_n)$ is the number of phonons in the box with energy $E_n$.

The size of the vibrating units sets a limit on the minimum wavelength since shorter wavelengths do not lead to new modes of the vibration. The smallest unit of crystalline solids is the unit cell. Thus the unit cell puts constrain on the minimum wavelength of the vibration as:

$$a \equiv \lambda_{min} \tag{18}$$

where a is the length of the unit cell. The minimum wavelength sets the maximum frequency $[\nu_{max}]$ and a highest mode number $[n_{max}]$ that an acoustic vibration can reach. The maximum frequency limits the vibrational energy, and the highest energy of acoustic vibrations is

$$E_{acoustic}^{max} = h\nu_{max} = \frac{h\upsilon_m}{\lambda_{min}}. \tag{19}$$

If all the atoms in the system possess the maximum acoustic vibrational energy then the maximum energy which can be reached by acoustic thermal vibration is:

$$U_{acoustic}^{max} = \sum_{n_x}\sum_{n_y}\sum_{n_z} nN_A E_{acoustic}^{max} = 3nN_A E_{acoustic}^{max} \tag{20}$$

Beyond this energy interacting or organized lattice vibration does not exist and the thermodynamic behavior of the system is described by independent lattice vibration. The internal energy of a system with independent thermal lattice vibration $[U_{thermal}]$ is given by Eq. (3). The temperature where the collective or acoustic vibration shifts to an independent thermal vibration is the Debye temperature, which can be defined as:

$$T = T_{Debye} \quad \text{when} \quad U_{acoustic}^{max} = U_{thermal} \tag{21}$$

Using the equality of the thermal internal energies Eq. (21) gives the Debye temperature as:

$$T_{Debye} = \frac{h\upsilon_m}{k_B \lambda_{min}}. \tag{22}$$

This expression is identical with the conventional definition given in Eq. (9). Most text books use the symbol $\theta_D$ for the Debye temperature since the definition is introduced in relation to vibrational frequencies. The minimum wavelength corresponding to the Debye temperature is then

$$\lambda_{min} = \frac{h\upsilon_m}{k_B \theta_D}. \tag{23}$$

This wavelength [Eq. (23)] should be equivalent with the length of the unit cell. This theoretical prediction is tested for monoatomic closest packing atomic arrangement.

### III. TESTING THE MODEL

The vibrational motion of crystals is very complex and the idealized Debye approach is valid only for monoatomic solids. Using the available acoustic velocity and Debye temperature data of elements with face centered cubic (fcc) (Al, Ag, Au, Cu, Ni, Pb, Pt) and hexagonal closed packing (hcp) (Mg, Ti, Zn) structure[18] the minimum wavelength of the phonon frequency at the Debye temperature is calculated. The calculated wavelength shows very good agreement with the size of the unit cell (Tab. 1). It is worth to notice that the phonon wavelength at the Debye temperature corresponds to the longer length of the unit cell (c) for hcp structures.

### IV. CONCLUSIONS

Physical model describing the Debye temperature is presented. The general formula for the Debye temperature is derived from fundamental quantum and thermodynamic assumptions. The wavelength of the phonon vibration at the Debye temperature is calculated for monoatomic systems with fcc and hcp structure. The calculated wavelengths of the cut off frequency correlates well to the unit cell parameters in agreement with the theoretical prediction.

[a]Electronic mail: jozsef.garai@fiu.edu

Table I.

Experimental values of the sound velocity, the Debye temperature the Unit cell parameters of the elements used for this investigation. The phonon wave length of the cut off frequency is calculated and compared to the unit cell parameters.

| Element | Crystal Structure | Transverse Sound Velocity [ms$^{-1}$] | Longitudinal Sound Velocity [ms$^{-1}$] | Mean Sound Velocity [ms$^{-1}$] | Debye Temperature [K] | Unit cell [Å] | Wavelength of the cut off frequency [Å] | Difference [%] |
|---|---|---|---|---|---|---|---|---|
| Ag | fcc | 3650 | 1610 | 1817.4 | 215 | 4.1[a, 19] | 4.06 | 0.8 |
| Al | fcc | 6420 | 3040 | 3420.4 | 394 | 4.1[a,*] | 4.17 | 17.0 |
| Au | fcc | 3240 | 1200 | 1362.2 | 170 | 4.1[a,*] | 3.85 | 6.8 |
| Cu | fcc | 4760 | 2325 | 2611.7 | 315 | 3.6[a, 19] | 3.98 | 10.2 |
| Ni | fcc | 6040 | 3000 | 3366.7 | 375 | 3.5[a, 19] | 4.31 | 22.4 |
| Pb | fcc | 1260 | 700 | 796.8 | 88 | 4.1[a, 19] | 4.34 | 7.3 |
| Pt | fcc | 3260 | 1730 | 1933.4 | 230 | 3.9[a,*] | 4.03 | 9.8 |
| Be | hcp | 12890 | 8880 | 9664.8 | 1031 | 3.58[c, 20] | 4.50 | 25.6 |
| Mg | hcp | 5770 | 3050 | 3409.4 | 330 | 5.21[c, 21] | 4.96 | 4.8 |
| Ti | hcp | 6070 | 3125 | 3499.4 | 380 | 4.68[c, 22] | 4.42 | 5.6 |
| Zn | hcp | 4210 | 2440 | 2707.9 | 234 | 4.95[c, 19] | 5.55 | 12.2 |

\* Equation of State of the substance was used to calculate the size of the unit cell at the Debye temperature.